\documentclass[twocolumn,showpacs,amsmath,amssymb,preprintnumbers,aip,jcp]{revtex4}
\usepackage{graphicx}
\usepackage{float}
\usepackage{dcolumn}
\usepackage{color}
\usepackage{bm}
\usepackage{gensymb}
\begin{document}

\title[M$_4$O$_4$ and M$_4$S$_4$ clusters]{ First principles study of electronic structure for cubane-like and ring-shaped structures of M$_4$O$_4$, M$_4$S$_4$ clusters (M = Mn, Fe, Co, Ni, Cu)}
\author{Soumendu Datta$^1$\footnote{Electronic mail: soumendu@bose.res.in} and Badiur Rahaman$^2$}
\affiliation{$^1$Department of Condensed Matter Physics and Material Sciences, S.N. Bose National Centre for Basic Sciences, JD Block, Sector-III, Salt Lake City, Kolkata 700 098, India \\
$^2$Department of Physics, Aliah University, IIA/27- Newtown, Kolkata 700 156, India}
\pacs{36.40.Cg, 73.22.-f,71.15.Mb}
\date{\today}

\begin{abstract}
Spin-polarized DFT has been used to perform a comparative study of the geometric structures and electronic properties for isolated M$_4$X$_4$ nano clusters between their two stable isomers - a planar rhombus-like 2D structure and a cubane-like 3D structure with M = Mn, Fe, Co, Ni, Cu ; X = O, S. These two structural patterns of the M$_4$X$_4$ clusters are commonly found as building blocks in several poly-nuclear transition metal complexes in inorganic chemistry. The effects of the van der Waals corrections to the physical properties have been considered in the electronic structure calculations employing the empirical Grimme's correction (DFT+D2). We report here an interesting trend in their relative structural stability - the isolated M$_4$O$_4$ clusters prefer to stabilize more in the planar structure, while the cubane-like 3D structure is more favorable for most of the isolated M$_4$S$_4$ clusters than their planar 2D counterparts. Our study reveals that this contrasting trend in the relative structural stability is expected to be driven by an interesting interplay between the $s$-$d$ and $p$-$d$ hybridization effects of the constituents' valence electrons.
\end{abstract}

\pacs{36.40.Cg, 73.22.-f,71.15.Mb}
\maketitle

\section{\label{sec:intro}Introduction}
After the discovery of graphene and  carbon fullerene based nano structures, the molecular clusters having cage-like as well as planar geometries have attracted much attention for constructing novel high-tech nano-materials. The special interest in such kind of structures, arises due to their unique electronic, optical, mechanical and chemical properties. Therefore, understanding of the mechanisms for their formation, has received immense interest during the last few decades. For the carbon-based nanomaterials, it is understood that the planar structure of the graphene sheet, is stabilized mainly by the $sp^2$ hybridization of the neighboring carbon atoms, while the spherical curvature of the cage-like structure of the carbon fullerene, produces an extra angular strain that allows a mixture of $sp^2$ and $sp^3$ hybridizations for bondings among the carbon atoms.\cite{reas1,reas2}Such fullerene-like or graphene-like structures have also been synthesized recently using nanoparticles of transition metal (TM) oxides or TM dichalcogenide compounds.\cite{tm_graphene1,tm_graphene2,tm_cage1,tm_cage2,tm_cage3,tm_cage4} Note that bulk TM oxides have already been studied extensively because of their wide range of applications from catalysis, organometallic, surface science to high temperature superconductivity, magnetic materials and so on. Furthermore, the TM oxides and chalcogenides being of semiconducting type in general, also lead to some obvious applications in sensors, electronics and solar cells. Nano structuring of these systems induces, in addition, an extreme sensitivity of their properties to the atomic arrangements and shapes of the systems due to quantum confinement effects of the structure. One of the most interesting facts about these nano clusters is that they exist in various substoichioetric compositions differing from their bulk behavior,\cite{asymetric} which renders them many unusual properties. Moreover, several oxide nano clusters exhibit some {\it magic} sizes and compositions which correspond to unusually high stability.\cite{magic} Due to the presence of $d$-electrons, the TM oxide clusters also show interesting magnetic properties. As for example, structural isomers of the oxide clusters possessing the same size and composition, are often found to have a rather different magnetic behavior.\cite{isomer}

With a thrust of gaining an atomistic understanding about the reactive properties in heterogeneous catalysis, many TM oxide nano clusters, in neutral or ionic forms, have recently been studied in gas phase by molecular beam experiments in combination  with mass spectrometry analysis.\cite{oxdcls1,oxdcls2} Such studies, indeed provide useful insights into the relationship between the geometric structures and the observed reactivity patterns. In the present study, our focus will be on the small nano clusters of the oxides and sulphides of 3$d$ late TM atoms. Each 3$d$ late TM atom is associated with a more than half-filled $d$-orbital. Therefore, both the hybridizations between the $3d$-orbital of the metal atoms with the $2p$ ($3p$)-orbital of the oxygen (sulfur) atoms in the corresponding oxide (sulphide) structure, and that between the valence 3$d$ and 4$s$ orbitals within the metal atoms itself, are expected to play an important role in deciding their many properties. While the nano clusters of TM oxides have been studied widely by both the experimental and theoretical works,\cite{oxdcls1,oxdcls2,oxdcls3,oxdcls4} the TM sulphide nano clusters have so far attracted less attention. It is interesting to follow-up that several molecular beam experiments on the isolated M$_m$O$_n$ oxide clusters of the 3$d$ late transition metal M atoms indicate a preference for the M:O ratio  as 1:1 particularly in case of the smaller cluster sizes.\cite{expt1,expt2,expt3,expt4}Attempts of searching for the structures of the smallest building blocks in the structures of such oxide nano particles, have also been offered recently by some first principles theoretical calculations.\cite{building_block1,building_block2}  We will concentrate here on the structure, electronic and magnetic properties of the isolated and isoatomic M$_4$O$_4$/M$_4$S$_4$ molecular building blocks with the metal atom M corresponding to one of the 3$d$ late transition metal atoms {\it i.e.} Mn, Fe, Co, Ni or Cu. Note that ring-shaped planar structures as well as cage-shaped structures have been predicted widely as the two competing stable isomers for the small clusters of most of the TM oxides.\cite{ref1,ref2} Our aim, in this work, will be, therefore, to perform a rigorous relative structural stability analysis of each M$_4$X$_4$ cluster between the two commonly found geometrical blocks, namely a 2D rhombus-like planar structure versus a 3D cubane-like structure. The special interest in these two structures, is mainly motivated by their relevance to multi-electron transfer centers in biological systems,\cite{bio1,bio2} their interesting magnetic and optical properties,\cite{mag,mag1,mag2,opt} as well as to their potential relevance to inorganic solids.\cite{inorg} Many poly-nuclear complexes containing cubane-type M$_4$O$_4$/M$_4$S$_4$ core units, have been studied extensively during the last decade. For examples, cubane-type Fe$_4$S$_4$ units exist in bacterial proteins and play  a variety of important roles in crucial cellular processes such as protein bio-synthesis and DNA replication.\cite{fe4s4a,fe4s4b} Catalytic activity by the cubane-type oxide clusters has also been explored in photo-system II for bio-inspired water splitting process.\cite{mn3cao4,mn4o4,co4o4} Interestingly, for understanding the catalytic mechanism in these systems, a ligand mediated $ring \Rightarrow cube$ structural rearrangement of the M$_4$X$_4$ molecular units, has recently been predicted as a crucial mechanism.\cite{catalytic}  Furthermore, the cubane structure allows a significant magnetic exchange between metal ions and therefore, the cubanes containing manganese atoms, in particular, have been studied extensively for their magnetic properties, specially in the search for single molecular magnets.\cite{yoo,miya,milios,beedle,karot} On the other hand, planar structure of the M$_4$X$_4$ units usually constitute building blocks for the structures of several layered perovskite oxides as well as cubic perovskite oxides, which have been studied extensively in recent time particularly for searching novel photo-catalysts of hydrogen production from water using the sun light.\cite{layer,cube,bg_variation}

In the present work using first principles density functional theory (DFT) based electronic structure calculations, we have performed a relative structural stability of the isolated M$_4$X$_4$ clusters between the two morphologies, as a first step towards understanding distinctly their role and the effects of the surrounding environment in the real systems of poly-nuclear transition metal complexes. Note that several M$_m$O$_n$ nano clusters have been studied in recent time for various metal M elements with varying $m$:$n$ ratios.\cite{oxdcls1_theory} Despite of the great progress achieved in this direction, a systematic study for the understanding of the general trends in structure, stability behavior and electronic properties among the oxide nano clusters of the whole 3$d$ late transition metal elements, is still lacking. Our aim here is to find out the trend in the relative structural stability along the oxide as well as sulphide nano clusters of all the 3$d$ late TM elements and obtain a better atomistic understanding about it. The hybridization of the valence orbitals of the constituent atoms has been quantified here in terms of a hybridization index parameter. Interestingly, our study here prevails that the relative stability of the isolated M$_4$X$_4$ systems between the two morphologies has been assessed basically by the predominance of either of these hybridization indexes. Present study reveals that the preference of the cubane-like 3D structures for the M$_4$S$_4$ clusters is governed by the dominant effect of the enhanced $p$-$d$ hybridization in this type of structure, while the effect of the enhanced $s$-$d$ hybridization plays the main role for the higher relative stability of the planar structure in case of the M$_4$O$_4$ clusters. We discuss our results of this study in the following in various sections. First, the computational details followed in this work, have been summarized in the Section \ref{methodology}. The optimized structures, stability and electronic properties, have been discussed in Section \ref{results}. The paper ends with a conclusion in Section \ref{concd}.

\section{\label{methodology} Computational Details} The calculations reported in this study, were performed using DFT within the framework of pseudo potential plane wave method, as implemented in the Vienna {\it ab-initio} Simulation Package (VASP).\cite{kresse2} We used the Projected Augmented Wave (PAW) pseudo potential \cite{blochl,kresse} coupled with the generalized gradient approximation (GGA) to the exchange correlation energy functional as formulated by Perdew, Burke and Ernzerhof (PBE).\cite{perdew} The 3$d$ as well as 4$s$ electrons for the TM metal atoms, the 2$s$ (3$s$) as well as 2$p$ (3$p$) electrons for the nonmetal oxygen (sulfur) atoms, were treated as the valence electrons. As the M$_4$X$_4$ clusters contain not only the metal atoms, but also the light nonmetal atoms, long range dispersion interactions could eventually be important and thus modifying the structures or the relative stability. To improve our description of the electronic properties of the M$_4$X$_4$ clusters, we also included the van der Waals interactions. We employed the empirical approach proposed by Grimme (DFT+D2),\cite{grimme} which describes the van der Waals interactions via a simple pair-wise force field. It has a lower computational cost and has currently been employed in the VASP code. In the DFT+D2 approach, the total energy, E$_{DFT+D2}$ is obtained by the sum of the conventional Kohn-Sham DFT energy (E$_{DFT}$) with the van der Waals dispersion correction (E$_{disp}$), {\it i.e}

\begin{equation}
E_{DFT+D2}=E_{DFT}+E_{disp},
\end{equation}
where
\begin{equation}
E_{disp}= -\frac{s_6}{2}\sum_i\sum_j\frac{C_6^{ij}}{R_{ij}^6}f_{dmp}(R_{ij}),
\end{equation}
where $i$ and $j$ run over the atoms in the unit cell. C$_6^{ij}$ represents the dispersion coefficient for the atom pair ($i$,$j$), $s_6$ is a global scaling factor that depends solely on the exchange-correlation functional ($s_6$=0.75 for PBE), R$_{ij}$ is the distance between the $i$ and $j$ atoms and $f_{dmp}$(R$_{ij}$) is a damped parameter employed in the DFT+D2 framework, as discussed in the Ref. \cite{grimme}.
 The wave functions were expanded in the plane wave basis set with the kinetic energy cut-off of 400 eV in all PBE and PBE+D2 calculations. The convergence of the energies with respect to the cut-off value were checked. For the cluster calculations, a simple cubic super-cell was used with periodic boundary conditions, where two neighboring clusters were kept separated by around 16 {\AA} vacuum space, which essentially makes the interaction between the cluster images negligible. Geometry optimizations were performed during the self-consistent calculations under the approximation of {\it collinear} magnetic orderings using the conjugate gradient and the quasi-Newtonian methods. The process of atomic relaxation was repeated until all the force components were less than a threshold value of 0.001 eV/{\AA} as well as the total energy difference of two consecutive relaxation steps was less than 10$^{-5}$ eV.  Reciprocal space integrations of the super-cell, were carried out at the $\Gamma$ point. The binding energy, E$_B$ of the optimal structures for each M$_4$X$_4$ cluster, is calculated with respect to the free atoms, as 
\begin{equation}
E_B(M_4X_4)=4E(X)+4E(M)-E(M_4X_4)
\end{equation}
where E(M$_4$X$_4$) is the total energy of the M$_4$X$_4$ cluster, while E(M) is the total energy of an isolated M atom. With this formula, E$_B$ is a +ve quantity and its more +ve value indicates higher stability.

\section{\label{results}Results and Discussions}

\begin{figure}
\rotatebox{0}{\includegraphics[height=6.8cm,keepaspectratio]{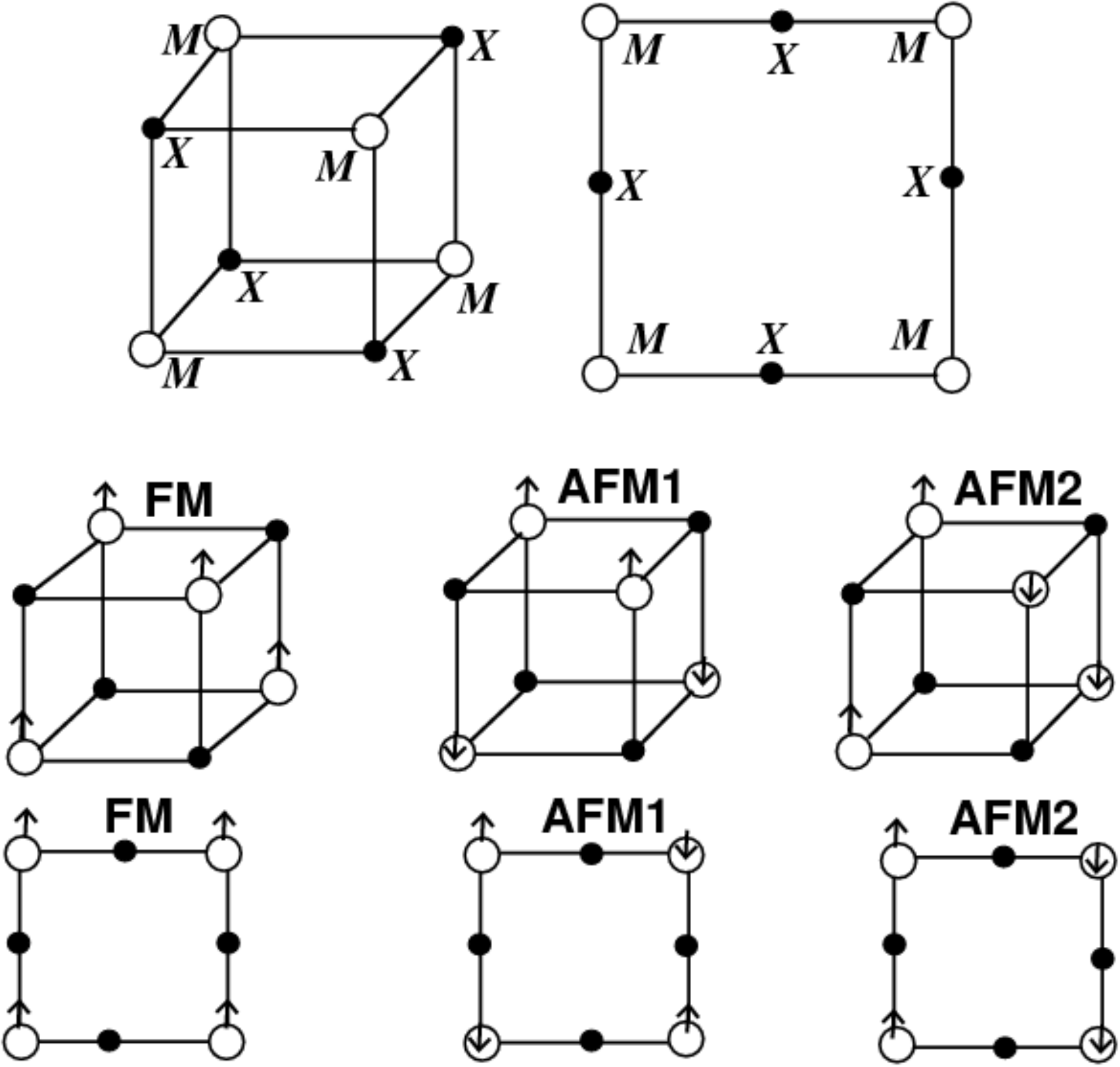}}
\caption{Two initial geometries (at top panels) of cubane-type 3D structure (left) and ring-shaped planar 2D structure (right) considered for each of the M$_4$X$_4$ clusters. The possible spin states with FM as well as AFM configurations within the moments of the transition metal M atoms in a M$_4$X$_4$ are shown for both the cubane (middle panels) and planar (bottom panels) structures, have been shown. The larger sized empty balls correspond to the transition metal atom and the smaller sized dots represent the nonmetal X atom in a M$_4$X$_4$ cluster.}
\label{structure}
\end{figure}

Theoretical investigations of the relative structural stability using DFT calculations, first requires the determination of the most  stable structures of the two geometries and the stabler one between them, is considered as the minimum energy structure (MES) out of the two geometries for each M$_4$X$_4$ cluster. For the nano clusters of TM oxides, the determination of the MES needs an exhaustive search down the potential energy surface (PES) as isomers of different magnetic characters are often found to lie in close energy separation and thereby, makes the task of determining the MES a cumbersome one. The primary structural patterns of the two most probable geometries for each M$_4$X$_4$ cluster, are shown in Fig. \ref{structure}. We have first taken these two geometries as the initial guessed structures for each M$_4$X$_4$ cluster and allowed them to relax for all possible spin configurations using {\it collinear} spin-polarized calculations. Instead of the global optimization, we allow the two structures to relax locally so that each structure undergoes for bond optimization only, while retaining their original shape. Therefore, the geometry relaxation for each system, has been done for all possible ferromagnetic (FM) as well as anti-ferromagnetic (AFM) configurations among the metal atoms of the 2D as well as 3D structures, also illustrated in the Fig. \ref{structure}. For the AFM couplings among the TM atoms, note that we have considered two spin configurations for each of the 2D and 3D structures. For an ideal cube, the two AFM configurations of the 3D cubane structure, appear identical. Consideration of distortion in the structure can, however, result different isomers. So, there are roughly  total six possible isomer/spin state combinations for each M$_4$X$_4$ cluster. Even for the FM configuration, each M$_4$X$_4$ cluster of both the ring and cage-shaped structures, were allowed to relax for all possible spin multiplicities to ensure the occurrence of intermediate spin state, if any, as the MES.

        After determining the optimized geometries, we have performed a comprehensive analysis on their relative stability as well as structural, magnetic and electronic properties.  The results of our DFT+D2 calculations have been discussed here. However, a comparison of the results of relative stability and magnetic coupling of the optimal 2D and 3D structures by the DFT+D2 as well as DFT without van der Waals correction, has been discussed later in this section. Table \ref{tab:bemag} shows the the details about the relative energies, metal-nonmetal nearest neighbor (NN) average bond lengths and total as well as average atom-centered magnetic moments of the most stable FM and AFM configurations for both the ring-shaped 2D as well as cubane-shaped 3D structures. The relative energy has been calculated with respective to the energy of the MES and its $+$ve values indicate that the corresponding isomers are higher in energy. To characterize the optimal structures, we have studied the average of the M$-$X NN bond lengths for each system. The average of the M$-$X bond lengths has been calculated on eight M$-$X NN bond lengths in case of the optimal 2D structures and twelve M$-$X NN bond lengths in case of the optimal 3D structures. The atom-centered magnetic moments have been calculated using the Mulliken population analysis of spin.\cite{muli} First focusing on the relative stability of the M$_4$O$_4$ clusters, it is seen from the Table \ref{tab:bemag} that the MES of each M$_4$O$_4$ cluster has the 2D ring-shaped structure. This is due to the fact that the 2D ring-shaped structures for the M$_4$O$_4$ clusters, are more stable than the respective most optimized 3D counterparts by an energy amount of 1.30 eV, 1.58 eV, 1.96 eV, 1.97 eV and 2.80 eV for M = Mn, Fe, Co, Ni and Cu respectively. Our calculations show that the MES of each M$_4$O$_4$ cluster, has AFM magnetic coupling for M = Mn, Fe, Co, Cu and an intermediate FM coupling with total magnetic moment of 4 $\mu_B$ for the MES of the Ni$_4$O$_4$ cluster. Also note that the most optimized 3D cubane structures have the AFM coupling for the Mn$_4$O$_4$, Fe$_4$O$_4$, Ni$_4$O$_4$ clusters, while the magnetic coupling is FM for the most stable 3D cubane structures of the Co$_4$O$_4$ and Cu$_4$O$_4$ clusters with total magnetic moments of 12 $\mu_B$ and 4 $\mu_B$ respectively. It is important to mention that our predictions of 2D ring-shaped geometry as the ground state structure for the M$_4$O$_4$ clusters and their preference for the AFM magnetic configuration, are mostly in accordance with the previous theoretical results.\cite{m4o4_old,fe4o4_old,co4o4_old,catalytic} Moreover, the presence of the oxygen atoms as a part in the MES of each M$_4$O$_4$ cluster, also plays significant role in deciding its structure and electronic properties, as the MES of the pure M$_4$ clusters would adopt a tetrahedron-like 3D structure for each of M = Mn, Fe, Co as well as Ni and favor a FM coupling with total magnetic moment of 20 $\mu_B$,\cite{Mn_pure} 12 $\mu_B$,\cite{Fe_pure} 10 $\mu_B$\cite{Co_pure} and 4 $\mu_B$\cite{isomer} respectively. However, a pure Cu$_4$ cluster is found to stabilize in planar structure with zero net magnetic moment.\cite{Cu_pure}

\begin{table*}[!t] 
\caption{\label{tab:bemag}Relative energy to the ground state 
($\triangle E=E-E_{\rm {ground \ state}}$), average NN bond lengths, total magnetic moments and distribution of magnetic moments of the optimal FM and AFM structures for both the 2D and 3D geometries. The $\triangle E$, total magnetic moments and magnitude of average magnetic moments within the parenthesis for the optimal FM structures, correspond to the respective values of the closely lying FM isomer having the same geometry.}
{\begin{tabular}{cccccccccccccccc} 
\hline
\hline
Cluster    & & Geometry & & Magnetic & & $\triangle E$ & & $\langle$M-X$\rangle$ &  & Total spin & &\multicolumn{2}{c} {Average spin ($\mu_B$/atom) at the site of} &  & \\ 
           &  &       & &coupling  &  &(eV)         & & ($\AA$)             &  & ($\mu_B$)  & &         &  & \\
\cline{13-16}
           &  &        & &          & &              &  &                   &   &            & & Metal {$\it i.e.$} $\langle$$\mu_M$$\rangle$  & Nonmetal {$\it i.e.$} $\langle$$\mu_X$$\rangle$  &  &       \\

\hline
Mn$_4$O$_4$& & Ring   &  & FM      &  &  0.562 [0.584]&   &      1.804      &    &  8 [10]   &  &  3.76 [4.07]        &  0.004 [0.085]  & & \\
           & &        &  & AFM     &  &  0.000        &   &      1.811      &    &  0        &  &  3.96               &  0.000          & & \\
           & & Cubane &  & FM      &  &  1.450 [2.123]&   &      1.992      &    &  10 [20]  &  &  4.06 [4.26]        &  0.105 [0.220]  & & \\
           & &        &  & AFM     &  &  1.299        &   &      1.986      &    &  0        &  &  4.01               &  0.070          & & \\
\hline
Fe$_4$O$_4$& & Ring   &  & FM      &  &  0.075 [0.377]&   &      1.780      &    &  8 [4]   &  &  3.15 [2.48]        &  0.200 [0.060]  & & \\
           & &        &  & AFM     &  &  0.000        &   &      1.770      &    &  0        &  &  3.03               &  0.130         & & \\
           & & Cubane &  & FM      &  &  1.626 [1.650]&   &      1.980      &    &  16 [6]  &  &  3.38 [3.15]        &  0.320 [0.060]  & & \\
           & &        &  & AFM     &  &  1.577        &   &      1.949      &    &  0        &  &  3.21               &  0.060          & & \\  
 \hline
 Co$_4$O$_4$& & Ring   &  & FM      &  &  0.226 [0.586]&   &     1.744      &    &  6 [8]   &  &  1.78 [1.62]        &  0.154 [0.230]  & & \\
           & &        &  & AFM     &  &  0.000        &   &      1.746      &    &  0        &  &  1.90               &  0.160          & & \\
           & & Cubane &  & FM      &  &  1.960 [2.213]&   &      1.942      &    &  12 [8]  &  &  2.37 [1.86]        &  0.420 [0.280]  & & \\
           & &        &  & AFM     &  &  2.184        &   &      1.922      &    &  0        &  &  2.16               &  0.190          & & \\
  \hline
 Ni$_4$O$_4$& & Ring   &  & FM      &  &  0.000 [0.035]&   &     1.724      &    &  4 [6]   &  &  1.30 [1.10]        &  0.200 [0.290]  & & \\
           & &        &  & AFM     &  &  0.257        &   &      1.720      &    &  0        &  &  1.08               &  0.000          & & \\
           & & Cubane &  & FM      &  &  2.048 [2.167]&   &      1.920      &    &  4 [2]  &  &  1.20 [0.90]        &  0.230 [0.120]  & & \\
           & &        &  & AFM     &  &  1.966        &   &      1.914      &    &  0        &  &  1.13               &  0.120          & & \\    
\hline
Cu$_4$O$_4$& & Ring   &  & FM      &  &  0.014 [0.468]&   &      1.745      &    &  0 [2]   &  &  0.00 [0.22]        &  0.00 [0.24]  & & \\
           & &        &  & AFM     &  &  0.000        &   &      1.740      &    &  0        &  &  0.15               &  0.10         & & \\
           & & Cubane &  & FM      &  &  2.795 [3.047]&   &      1.964      &    &  4 [2]  &  &  0.47 [0.23]        &  0.45 [0.23]  & & \\
           & &        &  & AFM     &  &  3.093        &   &      1.961      &    &  0        &  &  0.44               &  0.43          & & \\  
\hline
\hline
\hline
 Mn$_4$S$_4$& & Ring   &  & FM      &  &  2.070 [2.221]&   &     2.215      &    &  10 [18]   &  &  4.09 [4.15]        &  0.004 [0.04]  & & \\
           & &        &  & AFM     &  &  1.663        &   &      2.076      &    &  0        &  &  3.95               &  0.000          & & \\
           & & Cubane &  & FM      &  &  0.385 [1.235]&   &      2.399      &    &  2 [20]  &  &  3.50 [4.26]        &  0.040 [0.130]  & & \\
           & &        &  & AFM     &  &  0.000        &   &      2.338      &    &  0        &  &  3.90               &  0.040          & & \\    
\hline   
 Fe$_4$S$_4$& & Ring   &  & FM      &  &  3.034 [3.078]&   &     2.159      &    &  16 [10]   &  &  3.40 [3.00]        & 0.200 [0.13]  & & \\
           & &        &  & AFM     &  &  2.050        &   &      2.157      &    &  0        &  &  3.20               &  0.050          & & \\
           & & Cubane &  & FM      &  &  0.432 [0.440]&   &      2.212      &    &  16 [14]  &  &  3.27 [3.00]        &  0.240 [0.150]  & & \\
           & &        &  & AFM     &  &  0.000        &   &      2.264      &    &  0        &  &  2.92               &  0.060          & & \\    
\hline
 Co$_4$S$_4$& & Ring   &  & FM      &  &  2.063 [2.664]&   &     2.126      &    &  10 [12]   &  &  2.05 [2.31]        & 0.200 [0.32]  & & \\
           & &        &  & AFM     &  &  2.667        &   &      2.113      &    &  0        &  &  2.27               &  0.130          & & \\
           & & Cubane &  & FM      &  &  0.000 [0.228]&   &      2.169      &    &  6 [4]  &  &  1.24 [1.00]        &  0.100 [0.100]  & & \\
           & &        &  & AFM     &  &  0.344        &   &      2.168      &    &  0        &  &  0.99               &  0.030          & & \\    
\hline
 Ni$_4$S$_4$& & Ring   &  & FM      &  &  2.189 [2.775]&   &     2.100      &    &  8 [6]   &  &  1.28 [1.04]        & 0.370 [0.220]  & & \\
           & &        &  & AFM     &  &  2.880        &   &      2.067      &    &  0        &  &  0.90               &  0.001          & & \\
           & & Cubane &  & FM      &  &  0.000 [0.256]&   &      2.174      &    &  2 [4]  &  &  0.33 [0.60]        &  0.100 [0.220]  & & \\
           & &        &  & AFM     &  &  0.211        &   &      2.180      &    &  0        &  &  0.31               &  0.040          & & \\    
\hline
 Cu$_4$S$_4$& & Ring   &  & FM      &  &  0.167 [0.238]&   &     2.098      &    &  2 [4]   &  &  0.27 [0.31]        & 0.27 [0.41]  & & \\
           & &        &  & AFM     &  &  0.000        &   &      2.072      &    &  0        &  &  0.01               &  0.01          & & \\
           & & Cubane &  & FM      &  &  0.304 [0.320]&   &      2.269      &    &  2 [4]  &  &  0.20 [0.38]        &  0.18 [0.36]  & & \\
           & &        &  & AFM     &  &  0.356        &   &      2.271      &    &  0        &  &  0.27               &  0.28          & & \\ 
\hline
\hline
\end{tabular} }
\end{table*}

\begin{figure}
\rotatebox{0}{\includegraphics[height=11.9cm,keepaspectratio]{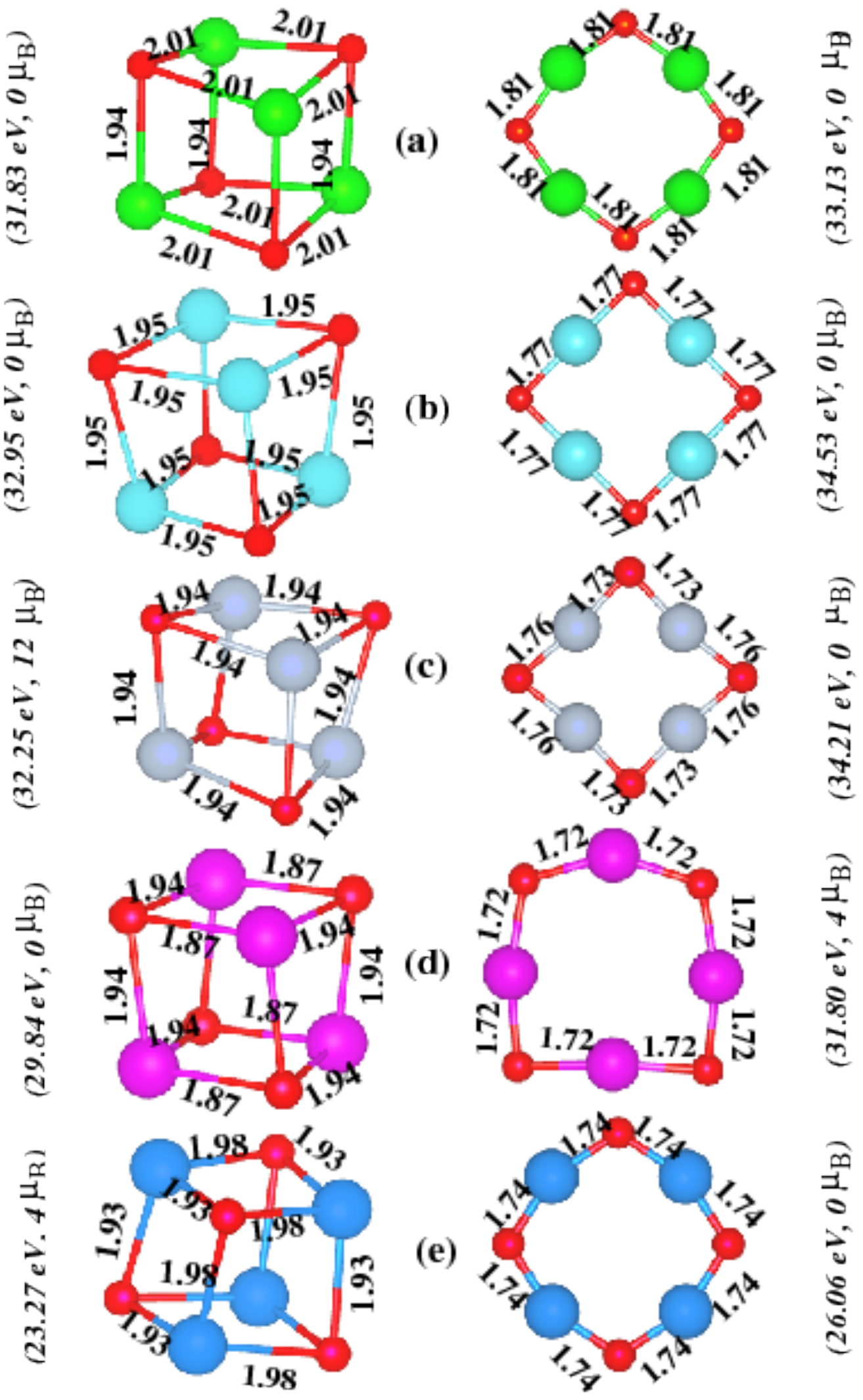}}
\caption{(Color online) The optimized structures of (a) Mn$_4$O$_4$, (b) Fe$_4$O$_4$, (c) Co$_4$O$_4$, (d) Ni$_4$O$_4$ and (e)Cu$_4$O$_4$ clusters in 3D cubane (left) and 2D ring (right) shaped configurations. The larger balls correspond to transition metal atom and the red colored smaller ball represents the Oxygen atom. The green, cyan, gray, magenta and blue colored larger balls correspond to Mn, Fe, Co, Ni and Cu atoms respectively. This choice of color for the metal atoms is also followed in the subsequent figures dealing with structure. The bond-lengths are given in Angstrom unit. The calculated binding energies and total magnetic moments of the optimal M$_4$O$_4$ clusters, are shown aside.}
\label{mes_m4o4}
\end{figure}

\begin{figure}
\rotatebox{0}{\includegraphics[height=11.9cm,keepaspectratio]{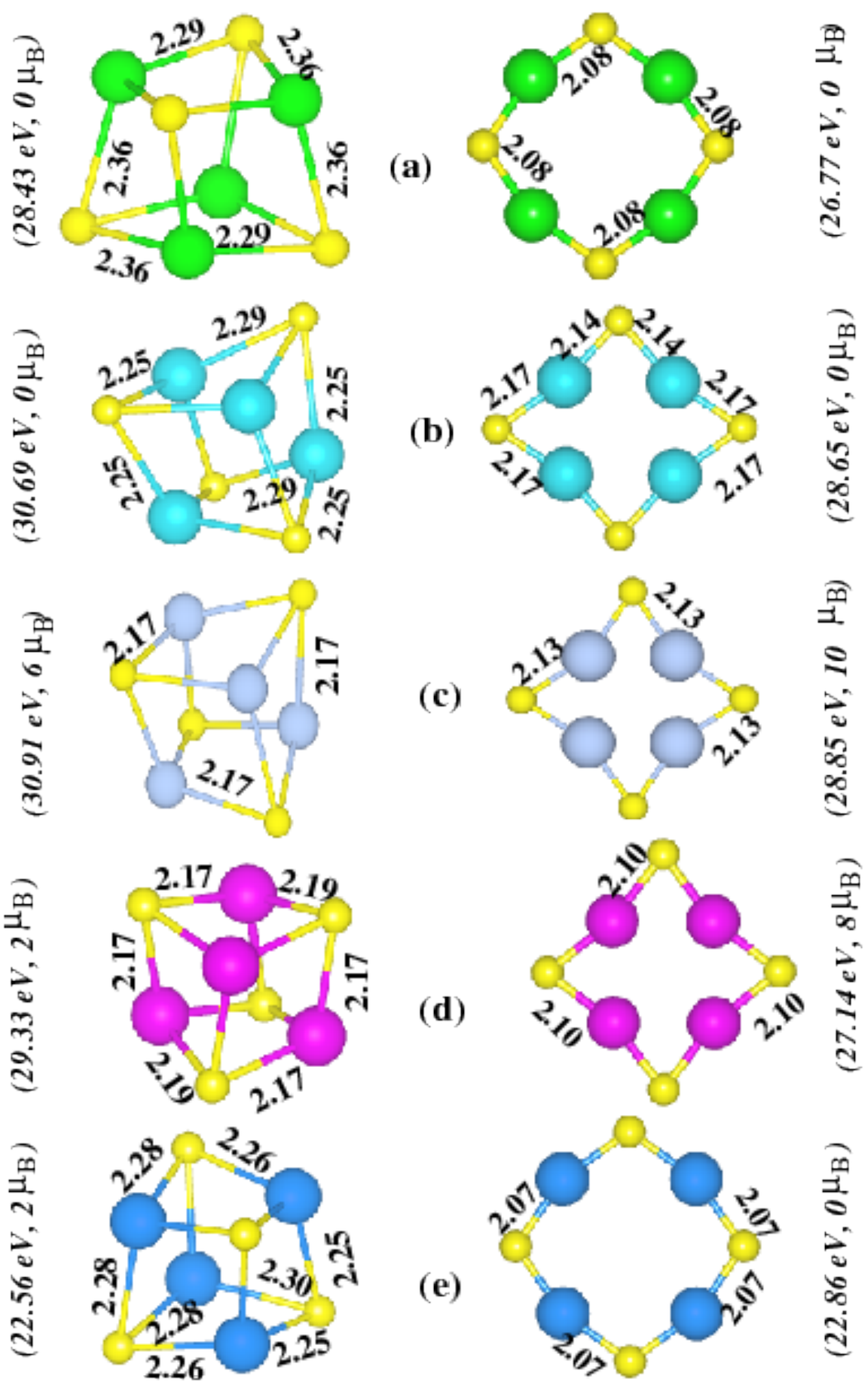}}
\caption{(Color online) The optimized structures of (a) Mn$_4$S$_4$, (b) Fe$_4$S$_4$, (c) Co$_4$S$_4$, (d) Ni$_4$S$_4$ and (e)Cu$_4$S$_4$ clusters in 3D cubane (left) and 2D ring (right) shaped configurations. The yellow colored smaller ball represents the Sulfur atom.  The bond-lengths are given in Angstrom unit. The calculated binding energies and total magnetic moments of the optimal M$_4$S$_4$ clusters, are shown aside.}
\label{mes_m4s4}
\end{figure}

On the other hand, the MES for each of the M$_4$S$_4$ clusters has 3D cubane-shaped  structure, except the case of the Cu$_4$S$_4$ cluster which rather continues to favor the 2D ring-shaped structure as the MES. In our calculations, the most optimal 3D structure for each of the Mn$_4$S$_4$, Fe$_4$S$_4$, Co$_4$S$_4$ and Ni$_4$S$_4$ clusters, is more stable than the most stable 2D counterpart for each system, by an energy amount of 1.66 eV, 2.05 eV, 2.06 eV and 2.19 eV respectively. We find that the 3D MESs of the Mn$_4$S$_4$ and Fe$_4$S$_4$ clusters, have AFM spin coupling among the metal atoms, while the 3D MESs for the Co$_4$S$_4$ and Ni$_4$S$_4$ clusters have FM spin coupling with the total magnetic moments of 6 $\mu_B$ and 2 $\mu_B$ respectively. Moreover, the most optimal ring-shaped 2D structures of the Mn$_4$S$_4$, Fe$_4$S$_4$ and Cu$_4$S$_4$ clusters have AFM coupling, while the most stable 2D ring-shaped structures of the Co$_4$S$_4$ and Ni$_4$S$_4$ clusters have FM coupling with the total magnetic moments of 10 $\mu_B$ and 8 $\mu_B$ respectively, which are higher than that of the total magnetic moment of their respective 3D MESs. Fig. \ref{mes_m4o4} and  Fig. \ref{mes_m4s4} show the most optimized 2D and 3D structures of the M$_4$O$_4$ and M$_4$S$_4$ clusters respectively. Our calculated binding energy, net magnetic moment and the NN bond lengths for each of the optimal structures, are also shown.

Next, to characterize the structures of the optimal M$_4$X$_4$ clusters, we concentrate on the variation of the average values of the M-X NN bond lengths. It is important to note that the average NN bond lengths of the optimal 2D and 3D structures as given in the Table \ref{tab:bemag} are, in fact exhibiting an interesting trend. This is in the sense that the $\langle$M-S$\rangle$ NN bond-lengths of the MESs of the M$_4$S$_4$ clusters, are generally higher than the $\langle$M-O$\rangle$ NN bond-lengths in cases of the 2D MESs for the M$_4$O$_4$ clusters with a common M. In addition, the Mulliken population analysis of atom-centered magnetic moments, whose average magnitudes are given in the Table \ref{tab:bemag}, also provides us a clear hint about the origin of the said magnetic exchange interactions of the constituent atoms in the optimal structures, as mentioned above. The spin population analysis for the MES of each M$_4$O$_4$ cluster, reveals that the moments at the oxygen sites are either zero or very small compared to that at the metal sites and thereby, can be regarded as nonmagnetic elements. Therefore, the moments of the four metal atoms and the very small moments associated with the four bridging oxygen atoms, interact with each other through {\it super-exchange} interaction which results in the observed AFM coupling within each other. In cases of the optimal FM coupling, on the other hand, we note that the hybridization between the metal $d$-orbital and nonmetal $p$-orbital plays an important role as it induces finite spin-polarization to the oxygen atoms and aligns them ferromagnetically with the moments of the metal atoms. Note that our interpretations of the AFM ground state in terms of the {\it super-exchange} interaction and the role of $p$-$d$ hybridization between the metal and nonmetal atoms in cases of the FM  ground states, also hold good in case of the MESs for the M$_4$S$_4$ clusters.

\begin{figure}
\rotatebox{0}{\includegraphics[height=5.9cm,keepaspectratio]{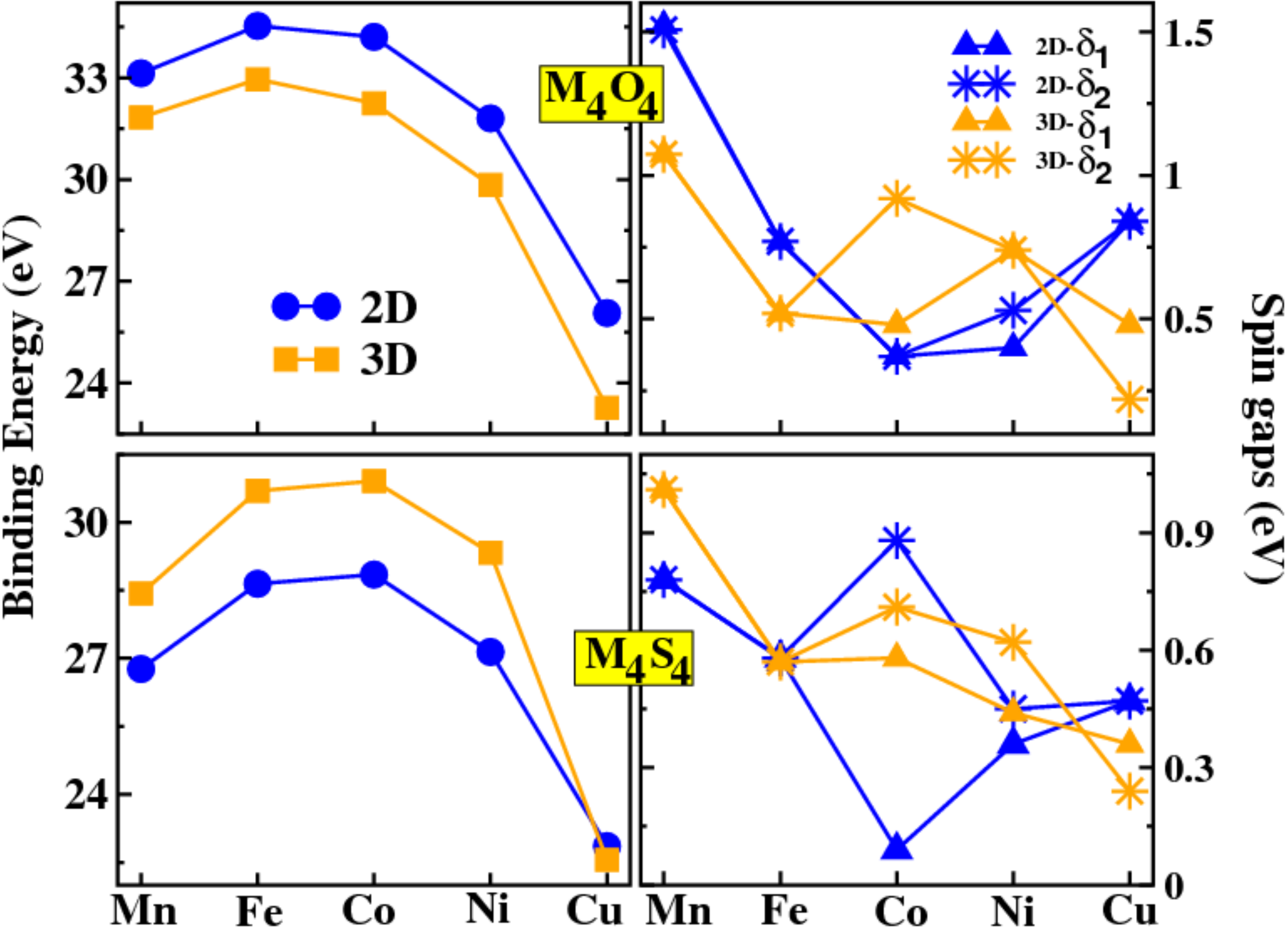}}
\caption{(Color online) Plot of binding energies in the left panels for the optimized M$_4$O$_4$ (top) and M$_4$S$_4$ (down) clusters in 3D (represented by light colored squares) and planar 2D (represented by blue colored balls) configurations. The right panels show the variation of spin gaps - $\delta_1$ (represented by triangles) and $\delta_2$ (represented by stars) in the optimized 2D and 3D structures of the M$_4$O$_4$ and M$_4$S$_4$ clusters. }
\label{gaps}
\end{figure}

Coming back to the issue of relative stability again, the most exciting point to note is that our analysis of energetics brings out a {\it unique} trend in the overall relative structural stability for the two classes of systems. Our first principles electronic structure calculations reveal that the M$_4$O$_4$ clusters are more stable in their optimized 2D geometry compared to its optimized 3D counterparts. In contrast to the M$_4$O$_4$ clusters, the M$_4$S$_4$ clusters prefer to stabilize more in the 3D cubane-like structures compared to their 2D planar counterparts. The Fig. \ref{gaps} shows the plot of our calculated binding energies of the most optimized 2D and 3D structures for both the M$_4$O$_4$ as well as the M$_4$S$_4$ clusters. The trend in relative stability as described above, is also reflected in the variation of their binding energies. This is in the sense that the 2D MESs of the M$_4$O$_4$ clusters have consistently higher binding energies compared to that of their respective optimized 3D counterparts. Likewise, the optimized 3D structures possess higher binding energies than their 2D counterparts for all the M$_4$S$_4$ clusters, except the Cu$_4$S$_4$ cluster, as seen in the trend of energetics for the M$_4$S$_4$ clusters in the previous section. We believe that this findings of the different kinds of variation of the relative structural stability of the two classes of the M$_4$X$_4$ clusters, with respect to a substitution of the four oxygen atoms by four sulfur atoms in going from the M$_4$O$_4$ cluster series to the M$_4$S$_4$ cluster series, is remarkable as microscopic chemistry appears to play the deciding role. Its understanding, therefore, could through light into the insights of the properties for the systems made up with these building blocks. It is important to mention that the trend in the relative structural stability is very robust being independent of the inclusion of van der Waals interaction. A comparison of our calculated $\triangle E$ - relative energy to the ground state, magnetic moment and the nature of the magnetic coupling for the minimum energy 2D and 3D structures using the DFT+D2 calculations as well as DFT calculations without van der Waals correction, has been given in the Table \ref{tab:comp}. It is clearly seen that the overall trend in the relative structural stability {\it i.e.} the preference of the 2D structure for the M$_4$O$_4$ clusters and the 3D structure for the M$_4$S$_4$ clusters, remains the same in the both cases. The only difference arises in the magnitude of the $\triangle$ E. Moreover, the nature of magnetic couplings as well as magnetic moment of the ground state structures and the preference for the 2D ring-shaped structure for the Cu$_4$S$_4$ cluster, remain mostly identical in the both cases.

\begin{table*}[!t] 
\caption{\label{tab:comp}Relative energy to the ground state ($\triangle E$), total magnetic moments and the nature of the magnetic coupling of the most optimal 2D and 3D geometries for each M$_4$X$_4$ cluster in calculations including van der Waals corrections (DFT+D2) and calculations without van der Waals corrections (DFT).}
{\begin{tabular}{ccccccccccccccccccc} 
\hline
\hline
Cluster    & & Geometry & &   \multicolumn{4}{c} {DFT+D2}         & &      &  &  & &   \multicolumn{4}{c} {DFT} &   & \\ 
\cline{5-9}\cline{13-17}
           &  &         & &   $\triangle E$  &  &  Total spin      & & Magnetic   &  &  & &   $\triangle E$  &  &  Total spin      & & Magnetic   &  &    \\
           &  &         & &      (eV)        &  &  ($\mu_B$)       & & coupling   &  &  & &      (eV)        &  &  ($\mu_B$)       & & coupling   &  &     \\
\hline
Mn$_4$O$_4$&  &   Ring  & &    0.000         &  &    0             & &  AFM       &  &  & &      0.000       &  &   0              & &  AFM       &  &   \\
           &  &  Cubane & &    1.299         &  &    0             & &  AFM       &  &  & &      1.470       &  &   0              & &  AFM       &  &   \\
\hline
Fe$_4$O$_4$&  &   Ring  & &    0.000         &  &    0             & &  AFM       &  &  & &      0.000       &  &   0              & &  AFM       &  &   \\
           &  &  Cubane & &    1.551         &  &    0             & &  AFM       &  &  & &      1.577       &  &   0              & &  AFM       &  &   \\
\hline
Co$_4$O$_4$&  &   Ring  & &    0.000         &  &    0             & &  AFM       &  &  & &      0.000       &  &   0              & &  AFM       &  &   \\
           &  &  Cubane & &    1.960         &  &    12            & &   FM       &  &  & &      1.906       &  &   12             & &   FM       &  &   \\
\hline
Ni$_4$O$_4$&  &   Ring  & &    0.000         &  &    4             & &  FM        &  &  & &      0.000       &  &   6              & &  FM       &  &   \\
           &  &  Cubane & &    1.966         &  &    0             & &  AFM       &  &  & &      1.896       &  &   0              & &  AFM       &  &   \\
\hline
Cu$_4$O$_4$&  &   Ring  & &    0.000         &  &    0             & &  AFM       &  &  & &      0.000       &  &   0              & &  AFM       &  &   \\
           &  &  Cubane & &    2.794         &  &    4             & &   FM       &  &  & &      2.956       &  &   4              & &   FM       &  &   \\

\hline
\hline
\hline

Mn$_4$S$_4$ & &  Ring   & &  1.663          &  &  0                &  &  AFM      &  &  & &       1.586      &  &   0              & &  AFM      &  &  \\   
            & &  Cubane & &  0.000          &  &  0                &  &  AFM      &  &  & &       0.000      &  &   0              & &  AFM      &  &   \\
\hline
Fe$_4$S$_4$ & &  Ring   & &  2.050          &  &  0                &  &  AFM      &  &  & &       2.254      &  &   0              & &  AFM      &  &  \\
            & &  Cubane & &  0.000          &  &  0                &  &  AFM      &  &  & &       0.000      &  &   0              & &  AFM      &  &  \\
\hline
Co$_4$S$_4$ & &   Ring  & &  2.063          &  &  10               &  &  FM       &  &  & &       2.758      &  &   12             &  &  FM      &  &  \\
            & &   Cubane & & 0.000          &  &  6                &  &  FM       &  &  & &       0.000      &  &    6             &  &  FM      &  &  \\
\hline
Ni$_4$S$_4$ & &   Ring  & & 2.189           &  &  8                &  &  FM       &  &  & &       2.871      &  &    6             &  &  FM      &  &  \\
            & &  Cubane & & 0.000           &  &  2                &  &  FM       &  &  & &       0.000      &  &    2             &  &  FM      &  &  \\
\hline
Cu$_4$S$_4$ & &  Ring   & & 0.000           &  &  0                &  & AFM       &  &  & &       0.000      &  &    0             &  &Nonmagnetic & & \\
            & &  Cubane & & 0.304           &  &  2                &  &  FM       &  &  & &       0.291      &  &    2             &  &  FM        & & \\
\hline
\hline
\end{tabular} }
\end{table*}

\begin{figure*}
\rotatebox{0}{\includegraphics[height=11.0cm,keepaspectratio]{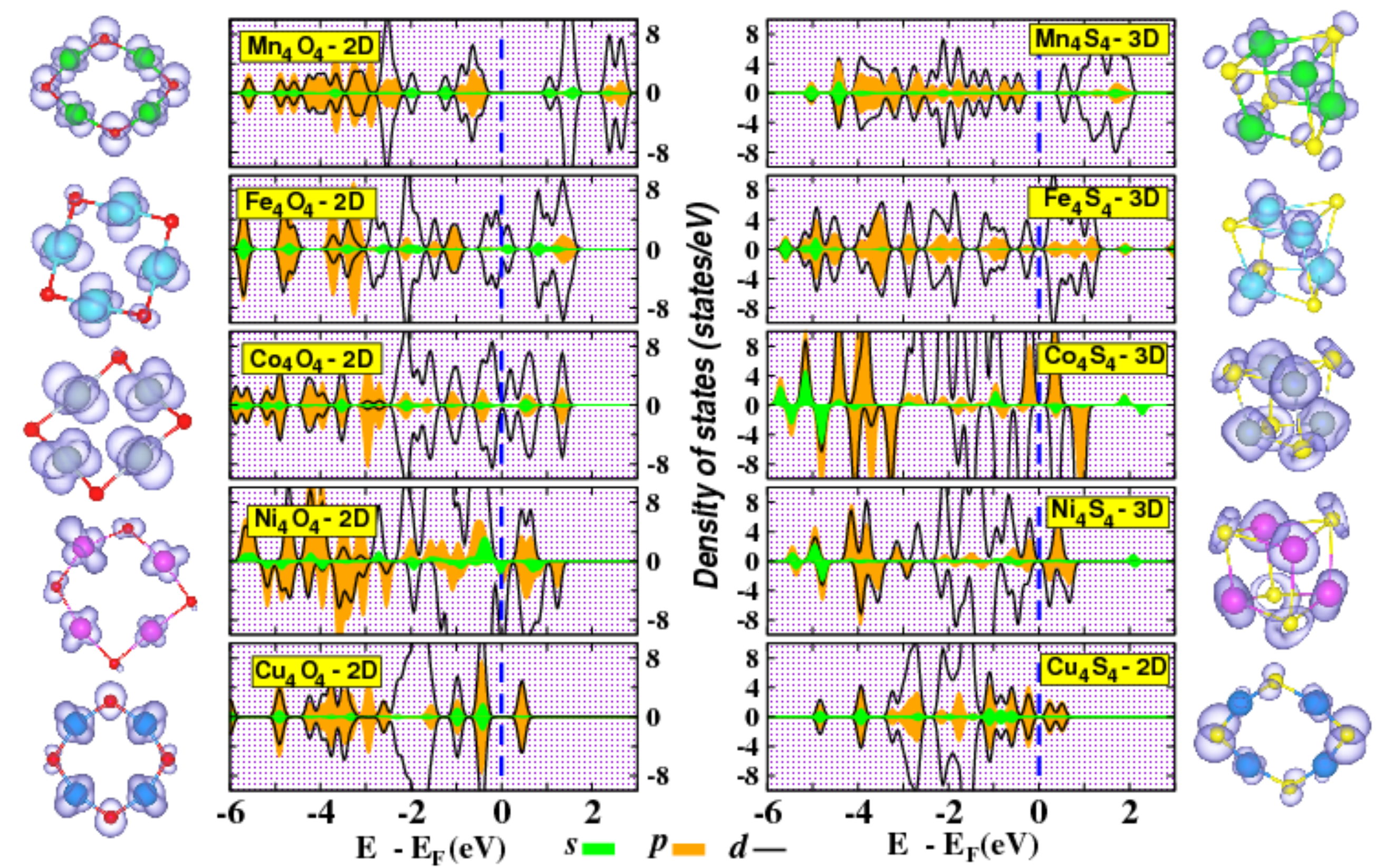}}
\caption{(Color online) $s$, $p$ and $d$ orbitals PDOS summed over all the constituent atoms for the MESs of M$_4$O$_4$ clusters (left panels) and M$_4$S$_4$ clusters (right panels). Gaussian smearing of 0.1 eV has been used. The Fermi energy along x-axis is fixed at zero indicated by the vertical dashed line. The HOMO isosurface for the MESs of the M$_4$O$_4$ and M$_4$S$_4$ clusters are shown aside. Isovale has been fixed at 0.005 e$^-$/\AA$^3$ for the plot of the orbital charge density.}
\label{dos}
\end{figure*}

To understand the stability behavior, we have studied the trend in the HOMO-LUMO spin-gaps for the optimal 2D and 3D structures for all the M$_4$X$_4$ clusters. The systems being magnetic, we have calculated two spin gaps - $\delta_1$ and $\delta_2$ as defined in our earlier work.\cite{spingap}  Our calculated values of the spin gaps are also plotted in the Fig. \ref{gaps}. It is seen that the $\delta_1$ and $\delta_2$ values  are positive for all the systems. Moreover, the 2D structures of the M$_4$O$_4$ clusters and 3D structures of the M$_4$S$_4$ clusters, mostly have higher values of $\delta_1$ and $\delta_2$, which is consistent with the fact of their relatively higher stability. While comparing the overall stability of the M$_4$O$_4$ clusters versus the M$_4$S$_4$ clusters, it is also seen from the Fig. \ref{gaps} that the M$_4$O$_4$ clusters have relatively higher binding energies compared to those of the M$_4$S$_4$ clusters for a common M element. In an attempt to understand the relatively lower stability and higher values of the NN bond-lengths for the optimal 3D M$_4$S$_4$ clusters, we have studied the structures and stability of the isolated M-O and M-S dimers. Our calculated binding energies and NN bond lengths for the optimal M-O dimers, are (5.44 eV, 1.64 \r{A}), (5.41 eV, 1.62 \r{A}), (5.37 eV, 1.61 \r{A}), (5.00 eV, 1.62 \r{A}) and (3.45 eV, 1.70 \r{A}) for M = Mn to Cu respectively. The calculated M$-$O bond lengths of the dimers which are much smaller than the respective bulk and also that of the M$_4$O$_4$ clusters, are in agreement with the experimental results.\cite{dimers} The corresponding values of the M-S dimers are (3.84 eV, 2.04 \r{A}), (4.35 eV, 2.00 \r{A}), (4.51 eV, 1.97 \r{A}), (4.39 eV, 1.96 \r{A}) and (3.25 eV, 2.03 \r{A}) which indicate that M-S dimers have relatively higher bond-length and  less stability compared to the corresponding M-O dimers. As each of the optimal 2D as well as 3D structures of the M$_4$X$_4$ clusters, consists of a number of the M$-$X dimers, the same trend in stability between the M$-$O and M$-$S dimers will also be reflected in the trend of relative stability between the M$_4$O$_4$ and M$_4$S$_4$ clusters. Therefore, the chemistry effect of replacing oxygen atoms by sulfur atoms in going from the M$_4$O$_4$ to the M$_4$S$_4$ clusters, plays a significant role in the predicted higher values of the $\langle$M-S$\rangle$ bond lengths as seen from the Table \ref{tab:bemag} and the lower stability for the optimal structures of the M$_4$S$_4$ clusters as seen from the  Fig. \ref{gaps}.

For better understanding about the trend in the relative structural stability of the M$_4$X$_4$ clusters between their 2D and 3D structures, we have studied the spin-polarized $s$, $p$ and $d$ orbital projected density of states (PDOS) summed over all the constituent atoms of the MESs of the M$_4$X$_4$ clusters as plotted in Fig. \ref{dos}. It is clearly seen that the $p$ and $d$ PDOS are dominating for both the M$_4$O$_4$ and M$_4$S$_4$ clusters. The contribution of the $s$ orbitals, though not significant, spreads over an extended energy ranges for the M$_4$O$_4$ clusters, while it is mostly localized to some particular energy intervals in case of the M$_4$S$_4$ clusters. It therefore, could result relatively larger $s$-$d$ hybridization for the M$_4$O$_4$ clusters compared to that of the M$_4$S$_4$ clusters. To examine the nature of splitting of PDOS near the Fermi energy, we have also plotted  in  Fig. \ref{dos} the HOMO orbital for the MES of each cluster. We note that the contributions of the $d_{xy}$ and $d_{x^2-y^2}$ orbitals of the transition metal atoms for the MESs of the 2D-M$_4$O$_4$ clusters are dominating. On the other hand, the dominating $d$-sub-orbitals for the transition metal atoms are mostly  $d_{xy}$, $d_{yz}$ and $d_{zx}$ for the MESs of the 3D-M$_4$S$_4$ clusters.

\begin{figure}
\vskip 0.8cm
\rotatebox{0}{\includegraphics[height=3.8cm,keepaspectratio]{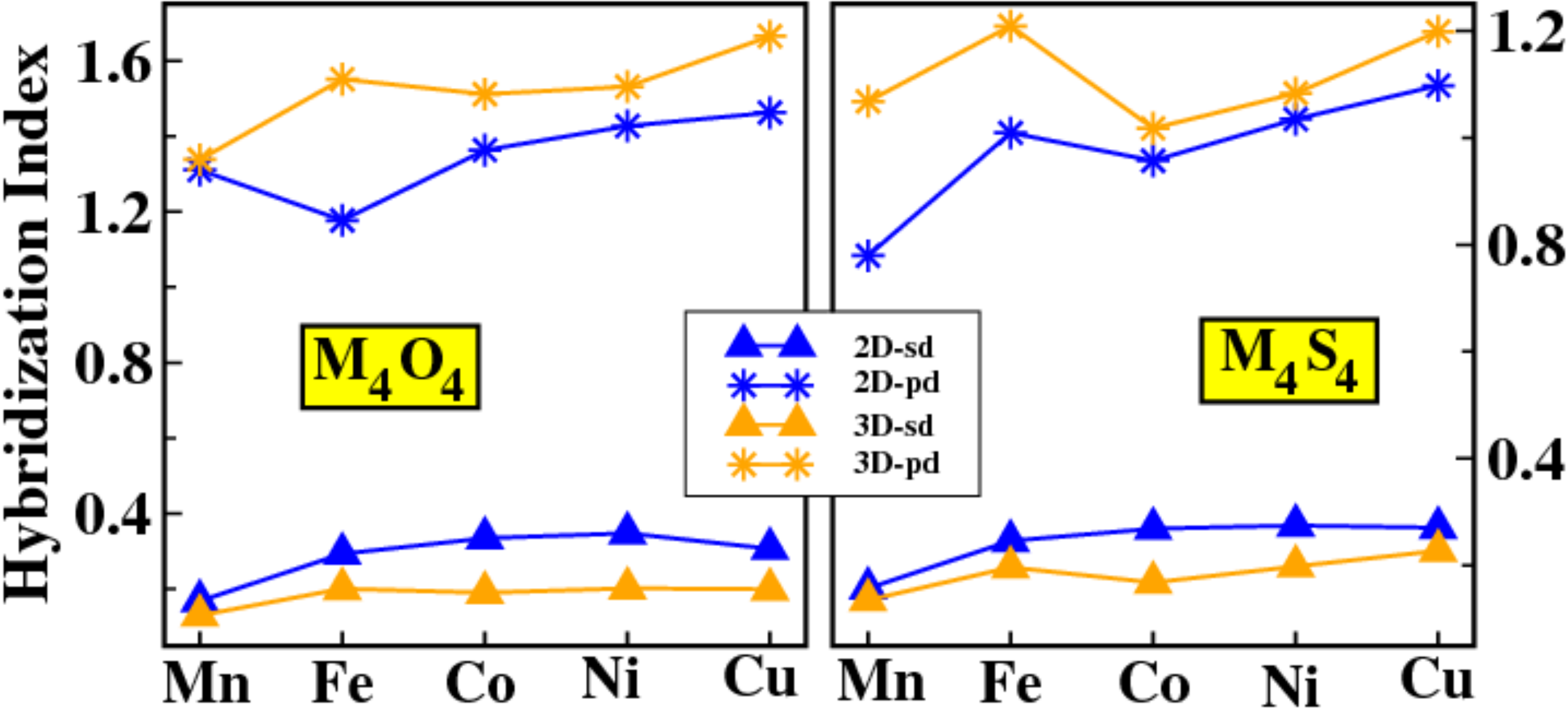}}
\caption{(Color online) Plot of the values of $s$-$d$ (represented by triangles) and $p$-$d$ (represented by stars) hybridization indices of the most stable 2D and 3D structures of both the M$_4$O$_4$ (left) as well as M$_4$S$_4$ (right) clusters. }
\label{stability}
\end{figure}

Finally, in order to quantify the orbital hybridizations, we have calculated $s$-$d$ as well as $p$-$d$ hybridization indexes for the most stable 2D and 3D structures of each M$_4$X$_4$ cluster. A hybridization index for a M$_4$X$_4$ cluster is generally defined by using the formula, $h_{kl}=\sum\limits_{I=1}^{8}\sum\limits_{i=1}^{occ}w_{i,k}^{(I)}w_{i,l}^{(I)}$ ; where $k$ and $l$ correspond to the orbital indices - $s$, $p$, $d$ and $w_{i,k}^{I}$ ($w_{i,l}^{I}$) is the projection of $i$-th Kohn-Sham orbital onto the $k$ ($l$) spherical harmonic centered at atom $I$, integrated over a sphere of specified radius. Note that the spin index is implicit in the summation. The role of such hybridization index for determining cluster morphology has been addressed previously.\cite{sd1,sd2,sd3} As the valance 4$s$ and 3$d$ orbitals of the metal atoms and the valence 2$s$/2$p$ (3$s$/3$p$) orbitals of the nonmetal oxygen (sulfur) atoms mainly participate in the hybridization, we have calculated both the $s$-$d$ as well as $p$-$d$ hybridization indexes. It is also important to mention at this point that these two hybridization indexes have counter effects with respect to controlling the stability behavior. This is in the sense that the overlap of the  transition metal atom $d$-orbitals (specially the $d_{xy}$, $d_{x^2-y^2}$ orbitals) with its $s$ orbitals in case of the systems with enhanced $s$-$d$ hybridization, confines mostly in the xy-plane and this would favor stabilization in a planar morphology. On the other hand, enhanced $p$-$d$ hybridization would involve hybridization of the nonmetal atom X-$p$ orbitals with the $d$ orbitals of the metal M atoms in all the three directions/planes and thereby, prones to favor a 3D geometry. Fig. \ref{stability} shows the plots of our calculated hybridization indexes for the most optimized 2D as well as 3D structures of all the M$_4$O$_4$ and M$_4$S$_4$ clusters. It is seen that irrespective of the 2D and 3D geometries, the $p$-$d$ hybridization index is larger than the $s$-$d$ hybridization index for both the optimal structures of each M$_4$X$_4$ cluster because of the significant overlapping of the valence $p$ orbital of X atoms with the valence $d$ orbital of the transition metal M atoms. Focusing on the relative magnitudes of the either hybridization index between the optimized 2D and 3D structures in the Fig. \ref{stability}, we note that the $p$-$d$ hybridization index is higher for the optimized 3D structures compared to that of their 2D counterparts. Conversely, the $s$-$d$ hybridization index is always higher for the optimized 2D structures than that of their 3D counterparts.  Therefore, though the overall variation in the magnitudes of the $s$-$d$ and $p$-$d$ hybridization indexes, follows the same trend between the optimal 2D and 3D structures for each system in either of the M$_4$X$_4$ series, it is their relative influences which come into play in determining the most stable structure. Combining the trends in the variation of stability as well as that of the hybridization indexes, it is, therefore, clearly indicating that the predicted stability behavior for both classes of the M$_4$X$_4$ cluster series, results microscopically from a delicate balance between the two types of the orbital hybridization. For the M$_4$O$_4$ clusters, the enhanced $s$-$d$ hybridization in favor of 2D structures, wins over the enhanced $p$-$d$ hybridization in favor of the 3D structures and overall stabilizes the 2D structures. Contrary to the case of M$_4$O$_4$ clusters, the effects of the enhanced $s$-$d$ hybridization in case of the optimized 2D structure of the M$_4$S$_4$ clusters is not significant as can be seen from their values in the Fig. \ref{stability} compared to those of the M$_4$O$_4$ clusters. As a result, it is not able to overcome the influence of its enhanced $p$-$d$ hybridization. Therefore, the enhanced $p$-$d$ hybridization index in favor of the optimized 3D structure appears as more dominant and stabilizes the 3D structures for most of the M$_4$S$_4$ clusters.

\section{\label{concd}Summery and Conclusions}
Using first principles DFT calculations, we have studied the structure, stability and electronic properties of the isolated M$_4$X$_4$ clusters between the two molecular patterns commonly found in inorganic chemistry - a cubane-like cage geometry as well as a ring-shaped planar geometry. As the X (O/S) atoms are characterized by the valence 2$s$ as well as 2$p$ orbitals and the transition metal atoms by the valence 4$s$ as well as 3$d$ orbitals, the influences of the orbital hybridization on the relative stability has been emphasized. We find that the relative structural stability results from an interesting interplay of $p$-$d$ {\it vis a vis} $s$-$d$ hybridizations. In case of the M$_4$O$_4$ clusters, the enhanced $s$-$d$ hybridization in favor of the 2D structures, wins over the enhanced $p$-$d$ hybridization in favor of the 3D counterparts. For the M$_4$S$_4$ clusters, however, the enhanced $p$-$d$ hybridization in favor of the 3D structure, topples the enhanced $s$-$d$ hybridization in favor of its 2D counterpart and overall stabilizes the cubane-like structure. This study will be useful for understanding as well as controlling the role of the environment surrounding the M$_4$X$_4$ molecular units in poly-nuclear transition metal complexes.
\section*{Acknowledgments}
We are grateful to Prof. T. Saha-Dasgupta for providing the computational facilities as well as for many simulating discussions. S. D. thanks Department of Science and Technology, India for support through INSPIRE Faculty Fellowship, Grant No. IFA12-PH-27.


\end{document}